# How to Price Shared Optimizations in the Cloud

Prasang Upadhyaya
Department of Computer
Science and Engineering
University of Washington
Seattle, WA, USA
prasang@cs.uw.edu

Magdalena Balazinska
Department of Computer
Science and Engineering
University of Washington
Seattle, WA, USA
magda@cs.uw.edu

Dan Suciu
Department of Computer
Science and Engineering
University of Washington
Seattle, WA, USA
suciu@cs.uw.edu

## ABSTRACT

Data-management-as-a-service systems are increasingly being used in collaborative settings, where multiple users access common datasets. Cloud providers have the choice to implement various optimizations, such as indexing or materialized views, to accelerate queries over these datasets. Each optimization carries a cost and may benefit multiple users. This creates a major challenge: how to select which optimizations to perform and how to share their cost among users. The problem is especially challenging when users are selfish and will only report their true values for different optimizations if doing so maximizes their utility.

In this paper, we present a new approach for selecting and pricing shared optimizations by using Mechanism Design. We first show how to apply the Shapley Value Mechanism to the simple case of selecting and pricing additive optimizations, assuming an offline game where all users access the service for the same time-period. Second, we extend the approach to online scenarios where users come and go. Finally, we consider the case of substitutive optimizations.

We show analytically that our mechanisms induce truthfulness and recover the optimization costs. We also show experimentally that our mechanisms yield higher utility than the state-of-the-art approach based on regret accumulation.

## 1. INTRODUCTION

Over the past several years, cloud computing has emerged as an important new paradigm for building and using software systems. Multiple vendors offer cloud computing infrastructures, platforms, and software systems including Amazon [3], Microsoft [10], Google [20], Salesforce [35], and others. As part of their services, cloud providers now offer data-management-in-the-cloud options ranging from highly-scalable systems with simplified query interfaces (*e.g.*, Windows Azure Storage [11], Amazon SimpleDB [9], Google App Engine Datastore [21]), to smaller-scale but fully relational systems (SQL Azure [26], Amazon RDS [6]), to data intensive scalable computing systems (Amazon Elastic MapReduce [4]), to highly-scalable unstructured data stores (Amazon S3 [8]), and to systems that focus on small-scale data integration (Google Fusion Tables [19]).

Existing data-management-as-a-service systems offer multiple options for users to trade-off price and performance, which we call generically *optimizations*. They include views and indexes (*e.g.*, users can create them in SQL Azure and Amazon RDS), but also the choice of physical location of data –which affects latency and price (*e.g.*, Amazon S3)– how data is partitioned (*e.g.*, Amazon SimpleDB data "domains"), and the degree of data replication (*e.g.*, Amazon S3 standard and reduced-redundancy storage). Cloud systems have an incentive to enable all the right optimizations, because this increases their customer's satisfaction and can also optimize the cloud's overall performance.

Today, data owners most commonly pay all costs associated with hosting and querying their data, whether by themselves or by others. Data owners also choose, when possible, the optimizations that should be applied to their data. However, there is a growing trend toward letting users collaborate with each other by sharing data and splitting data access costs. For example, in the Amazon S3 storage service, users can currently share their data with select other users, with each user paying his or her own data access charges [7].

The combination of data sharing and optimizations creates a major challenge: how to *select the optimizations to implement and how to price them when one optimization can benefit multiple users.* Implementing these optimizations imposes a cost on the cloud that needs to be recovered: resources spent on implementing and maintaining optimizations are resources that cannot be sold for query processing.

A recently-proposed approach by Kantere, Dash, *et al.*, [16, 22] addresses this problem by asking users to indicate their willingness to pay for different query performance values, observing the query workload, and deciding on the optimizations to implement based on optimizations that would have been helpful in the past (*i.e.*, based on *regret*). The cost of the implemented optimizations is amortized over the future queries that make use of them. This approach, however, has two key limitations as we show in Section 8. First, it assumes that users in the cloud will truthfully reveal their valuations. In practice, users will try to game the system if doing so improves their own utility. Other collaborative systems like peer-to-peer networks experience widespread gaming [2] that can degrade system performance [17], and incentives to reduce gaming are core components of modern peer-to-peer clients [15]. Second, this approach does not guarantee that the cost of an optimization will be recovered.

Given these two observations, we develop a new approach to select and price optimizations in the cloud based on Mechanism Design [31, 33]. Mechanism Design is an area of game theory whose goal is to choose a game structure and payment scheme such as to obtain the best possible outcome to an optimization problem in spite of *selfish players having to provide some input to the optimization*. Our goal is to enable the cloud to find the best configuration of optimizations. For this, it needs users (*i.e.*, selfish players) to reveal their valuations for these optimizations.

The most closely related approaches from the Mechanism Design literature are cost-sharing mechanisms [27]. Given a service with some cost, these mechanisms decide what users to service and how much the users should pay for the service. We show how to easily adapt this technique from the game theory community to the simplest problem of pricing a single optimization when all users access the system for a single time-period (*i.e.*, offline games).

The problem of pricing optimizations in the cloud, however, raises two additional challenges. First, in the cloud, users change their workloads as well as join and leave the system at any time. Such dynamism complicates the problem because the choice and price of optimizations must vary over time (*i.e.*, we need an online mechanism), and users now have new ways of gaming the system: they can lie about the time when they need an optimization and they can emulate multiple users. Second, multiple optimizations are available in the cloud and the value that a user derives from these optimizations can be given by a complex function. In particular, in this paper, we consider *additive*, or independent, optimizations and *substitutive*, or equivalent, optimizations.

We seek the following standard properties for our mechanisms. First, we want the mechanisms to be *truthful*, also known as *strategy-proof* [31], which means that every player should have an incentive to reveal her true value obtained from each optimization. The approach by Dash, Kantere *et al.* [16, 22] mentioned above is not truthful as we discuss in Section 8: users can benefit from lying about their value for an optimization. We also want online mechanisms to be resilient to multiple identities and to misrepresentation of the time when a user needs an optimization. Second, we want the mechanisms to be *cost-recovering*, which means that the cloud should not lose money from performing the optimizations. In the approach by Dash, Kantere, *et al.* [16, 22], the cloud first decides to implement an optimization and then the cost is amortized over the future queries that use it. Cost-recovery is thus not guaranteed. Finally, we want the mechanisms to be *efficient*, also known as *value-maximizing* [31], which means that we want it to maximize the total social utility of the system *i.e.*, the sum of user values minus the cost of the implemented optimizations. For example, if several users could benefit from an expensive optimization that none of them can afford to pay for individually, then the cloud should perform the optimization and divide the cost among the users.

In summary, we make the following four contributions:

We first show how the problem of pricing optimizations maps onto a cost-recovery mechanism design problem (Section 3). We also show how the Shapley Value Mechanism [27], which is known to be both cost-recovering and truthful, solves the problem of pricing a single optimization.

We propose a direct extension of the mechanism to the case of additive optimizations in an *offline* scenario, where all users access the system for the same time-period. We call this basic mechanism $Add^{Off}$ *Mechanism* (Section 4).

Second, we present a novel mechanism for the online scenario where users come and go, called the $Add^{On}$ *Mechanism*. It turns out to be much more difficult to design mechanisms for the online setting: algorithms that are truthful or cost-recovering in the static setting cease to be so in the dynamic setting (see [31, p. 412]). We prove our new mechanism to be both cost-recovering and truthful in the dynamic setting (Section 5).

Third, we extend both the $Add^{Off}$ Mechanism and the $Add^{On}$ Mechanism to the case where optimizations are inter-dependent, or substitutive. We call these mechanisms $Subst^{Off}$ *Mechanism* and $Subst^{On}$ *Mechanism* and prove them truthful (assuming users do not know other users' valuations) and cost-recovering (Section 6).

It has been proven before that achieving both truthfulness and cost-recovery, in the face of selfish agents, comes at the expense of total utility [27]. We experimentally compare our mechanisms against the state-of-the art approach based on regret accumulation [16] and show that our mechanisms produce up to a 3× higher utility and provide the same utility for ranges of optimization costs up to 12.5× higher than the state-of-the-art approach in addition to handling selfish users and ensuring that the cloud recovers all costs.

## 2. MOTIVATING USE-CASE

An important component of the astronomy research conducted by our colleagues in the astronomy department at the University of Washington involves large universe simulations [23], where the universe is modeled as a set of particles, which include dark matter, gas, and stars. All particles are points in a 3D space with properties that include position, mass, and velocity. Every few simulation time steps, the simulator outputs a snapshot of the state of the universe capturing all properties of all particles at the time of the snapshot. State of the art simulations (*e.g.*, Springel et al. [37]) use over 10 billion particles producing a dataset of over 200 GB per snapshot.

For each snapshot, astronomers first run a clustering algorithm to detect clusters, called *halos*. Some halos correspond to galaxies. Studying the evolution of these halos over time is a major component of their research. Different astronomers research different types of halos. In particular, our colleague indicated that: "There are in general three or four different halo mass ranges that different people focus on: high mass which corresponds to a cluster, Milky Way mass, slightly less than Milky Way mass and low mass/dwarf galaxies. [...] For example, I've been looking for Milky Way Mass galaxies, but another person in our group might be interested in the same sort of galaxies, but at a lower mass range. [The simulation] also helps us identify what environment a given halo forms in – one person might be interested in a Milky Way mass galaxy that forms in relative isolation, another person might be interested in finding a Milky Way mass galaxy that forms near many other galaxies (a rich, cluster-like environment)." [25]. Additionally, different scientists focus on different particle types and on the simulation time steps that correspond to interesting time-periods in the evolution of the halos that they study [25]. Thus, different users may need different optimizations (indexes and materialized views for this use-case), and the challenge is to decide which ones to implement, and who pays for them.

In Section 7.2, we evaluate our mechanisms on real data and queries (optimized using materialized views) from this use-case. Since different scientists query different parts of the data, they benefit from different materialized views.

## 3. A MECHANISM DESIGN PROBLEM

In this section, we show how to model the problem of selecting and pricing optimizations in the cloud as a *mechanism design* [31] problem. We further show that our problem requires a type of mechanism called *cost-sharing mechanism*. In this paper, we assume that every optimization is binary, *i.e.*, the cloud either implements it or not. We do not consider continuous optimizations (*e.g.*, degree of replication).

We consider a set of users, $I = \{1, \ldots, m\}$, who are using a cloud service provider (*a.k.a.*, *cloud*) to access and query several datasets. Any user can potentially access any dataset. Let $J = \{1, \ldots, n\}$ be the set of all potential optimizations that the cloud offers for these datasets. For example, $j$ may represent an index; or the fact that a dataset is replicated in another data center; or may be an expensive fuzzy join between two popular public datasets, which is precomputed and stored as a materialized view. Upon deciding to do an optimization $j$, the cloud may restrict access to $j$ to only certain users; a *grant pair* $(i, j)$ indicates that user $i$ has been granted permission to use the optimization $j$. While grant permissions artificially prevent a user from accessing an optimization, this restriction is required to ensure that users reveal their true value for an optimization and pay accordingly. A *configuration*, also called *alternative*, is a set of optimizations $j$ and a set of grant pairs[1] $(i, j)$. We denote an alternative with $a$ and the set of all possible alternatives with $A$. We also denote $S_j = \{i \mid (i, j) \in a\}$ to be the users who get access to the optimization $j$ in alternative $a$.

The goal of the mechanism will be to select a configuration $a \in A$. The decision will be based on the optimization costs and their values to users, which will determine the users' willingness to pay for various optimizations.

**Values to Users.** Each user $i$ obtains a certain value $v_{ij} \geq 0$ from each optimization $j$: *e.g.*, monetary savings obtained from faster execution or the ability to do a more complex data analysis. When multiple optimizations are performed, the total value to a user is given by $V_i(a) \geq 0$, and is obtained by aggregating the values $v_{ij}$ for all grant pairs $(i, j) \in a$. In this and the following two sections, we consider *additive optimizations*, where the value is given by:

$$V_i(a) = \sum_{(i,j) \in a} v_{ij} \quad \geq 0 \tag{1}$$

We consider *substitutive optimizations* in Section 6.

An important assumption in mechanism design is that users try to lie about their true values: when asked for their value $v_{ij}$, user $i$ replies with a bid $b_{ij}$, where $b_{ij}$ may be different from $v_{ij}$. In the case of an additive value function, we denote $B_i(a) = \sum_{(i,j) \in a} b_{ij}$, where $B_i(a)$ is user $i$'s *bid* about her value $V_i(a)$.

**Cost to the Cloud.** For each implemented optimization $j \in J$, the cloud incurs an optimization cost $C_j > 0$, which includes the initial cost of implementing the optimization (*e.g.*, building an index) and any possible maintenance costs (*e.g.*, updating the index) for the duration of the service. This cost is an opportunity cost: the resources used to perform the optimization cannot be sold to other users. The cost of an alternative $a$ is then given by:

$$C(a) = \sum_{j \in a} C_j \tag{2}$$

Even if each cost $C_j$ is small, the combined cost $C(a)$ may be large since the number of potential optimizations is large.

**Payments.** Once an outcome $a$ is determined, each user $i$ who is granted access to an optimization $j$ must pay some amount $p_{ij}$. This payment is called the user's *cost-share*, and is determined based on all users' bids[2], $(b_{ij})_{i=1,m;j=1,n}$. If $P_i = \sum_j p_{ij}$ is the total payment for user $i$, her *utility* is defined as $U_i(a) = V_i(a) - P_i$. A standard assumption in Mechanism Design is that users are "utility maximizers", *i.e.*, they bid to maximize their utility [31, 33].

**Cost-Sharing Mechanism Design Problem.** After collecting all bids, the mechanism chooses an outcome $a_0 \in A$ that optimizes some global value function. In the case of cloud-based optimizations, we will aim to optimize the *total social utility* ("total utility" for short): the outcome's total value (Eq. 1) minus the outcome's cost (Eq. 2). Formally, the mechanism chooses the following outcome $a_0$:

$$a_0 = \arg\max_{a \in A} \left( \sum_{i \in I} B_i(a) - C(a) \right) \tag{3}$$

Such a mechanism is called *efficient* [27]. Note that the mechanism does not know the true values $V_i(a)$, but uses the bids $B_i(a)$ instead. The goal of *mechanism design* is to define the payment functions $p_{ij}$ so that all users have an incentive to bid their true values $B_i = V_i$. A mechanism is called *strategy-proof* [31, 33], or *truthful*, if no user can improve her utility $U_i(a)$ by bidding untruthfully, *i.e.*, with $B_i \neq V_i$. Truthful mechanisms are highly desirable, because when users reveal their true values, the mechanism is in a better position to select the optimal alternative.

Another desired property for cost-sharing mechanisms is to be *cost-recovering*, *i.e.*, to only pick outcomes $a_0$ so that:

$$C(a_0) \leq \sum_i P_i \tag{4}$$

EXAMPLE 1. *Consider a naïve mechanism: The cloud collects all bids $b_{ij}$; if $c_j \leq \sum_i b_{ij}$, it performs the optimization $j$ and asks each user to pay $b_{ij}$ ($p_{ij} = b_{ij}$). Clearly it is cost-recovering. However, it is not truthful: a user $i$ can lie and declare a much lower value $b_{ij} \ll v_{ij}$, hoping that the optimization would be performed anyway and she would end up paying much less than her true value. The challenge in designing any mechanism is to ensure its truthfulness.*

Formally, a mechanism is defined as follows:

DEFINITION 1. *A mechanism $(f, P_1, \cdots, P_m)$ consists of a function $f : (\mathbb{R}^A)^m \to A$ (called* social choice function*) and a vector of payment functions $P_1, \cdots, P_m$, where $P_i : (\mathbb{R}^A)^m \to \mathbb{R}$ is the amount that user $i$ pays.*

---
[1]We assume that, if an alternative contains a grant pair $(i, j)$, then it also contains the optimization $j$.

[2]This is a very important point: the payment depends not only on the outcome $a$, but on all bids. For *e.g.*, in the *second bidders' auction*, the winner's payment is the second highest bid [33].

| Symbol | Description |
|---|---|
| **i, j, t, a** | Index for users, optimizations, time-slots and outcomes. |
| **I, J, T, A** | Sets of users, optimizations, time-slots and outcomes. |
| $\mathbf{S_j(t)}$ | Users serviced by optimization $j$ at time $t$. |
| $\mathbf{CS_j(t)}$ | All users serviced by optimization $j$ up until time $t$. |
| $\mathbf{v_{ij}(t)}$ | User $i$'s true (private) value for optimization $j$ at time $t$. |
| $\mathbf{b_{ij}(t)}$ | User $i$'s stated value for optimization $j$ at time $t$. |
| $\mathbf{B_i}$ | $\mathbf{B_i} = (b_{ij})_{i=1,m; j=1,n}$. |
| $\mathbf{V_i(a)}$ | User $i$'s total, true (private) value for outcome $a$. |
| $\mathbf{B_i(a)}$ | User $i$'s total, stated (public) value for outcome $a$. |
| $\mathbf{p_{ij}}$ | User $i$'s payment for optimization $j$. |
| $\mathbf{P_i}$ | User $i$'s total payment. |
| $\mathbf{U_i(a)}$ | User $i$'s utility for outcome $a$. |
| $\mathbf{C(a), C_j}$ | Outcome $a$'s cost, and optimization $j$'s cost, respectively. |
| $\mathbf{s_i}$ | Slot when user $i$ enters the system. |
| $\mathbf{e_i}$ | Slot when user $i$ pays and leaves the system. |

Table 1: Symbol Table. For symbols with the argument time $t$, we drop $t$ for offline mechanisms.

The mechanism works as follows. After collecting bids $B_1, \ldots, B_m$ from all users[3], it chooses the alternative $a = f(B_1, \ldots, B_m)$ where each user $i$ must pay $P_i(B_1, \ldots, B_m)$.

While we would like to design mechanisms that maximize the total utility (Eq.(3)), it is a proven result that one cannot achieve cost-recovery (*a.k.a.* budget-balance), truthfulness and efficiency [27] simultaneously. In our setting, we ensure only truthfulness and cost-recovery (Eq.(4)) at the expense of some efficiency loss. Indeed, if the cloud cannot recover its cost, it will not implement the loss-making optimization.

## 4. A MECHANISM FOR STATIC COLLABORATIONS

We now show how to use the Shapley Value Mechanism [27], which has many desirable properties, to solve the problem of selecting and pricing *additive* optimizations for *one time-slot* (*i.e.*, offline games). We extend it to online settings, where users come and go across multiple time-slots in Section 5 and to substitutive optimizations in Section 6. For ease of reference, we summarize the notations used in this paper in Table 1.

### 4.1 Background: Shapley Value Mechanism

We start by reviewing the Shapley Value Mechanism [27], shown in Mechanism 1. Fix a single optimization $j$, let $C_j$ be its cost and $b_{1j}, \ldots, b_{mj}$ the users' bids for this optimization. Mechanism 1 determines whether to perform the optimization or not, and, computes the set of serviced users $S_j \subseteq \{1, \ldots, m\}$, and how much they have to pay, $p_{ij}$. Intuitively, it finds the minimum price $p$ to charge to each user who bid more than $p$ such that the total payment is at least $C_j$. It starts by setting $S_j$ to the set of all users, and divides the cost $C_j$ evenly among them: $p = C_j/|S_j|$. If $p$ is larger than a user's bid $b_{ij}$, she is removed from $S_j$ and a new price is recomputed by dividing the cost evenly among the remaining users. As a result, the cost per user, $C_j/|S_j|$, may increase and additional users may need to be removed from the set $S_j$. The process continues until either no users remain or no further users need to be removed from $S_j$. Each serviced user $i \in S_j$ pays the same amount $p_{ij} = C_j/|S_j|$; each non-serviced user $i \notin S_j$ pays nothing, *i.e.*, $p_{ij} = 0$. If $S_j = \emptyset$, no subset of users has bid enough to pay for the optimization, and it is not implemented at all. It is obvious that this mechanism is cost-recovering, since $\sum_{i \in S_j} p_{ij} = C_j$. The mechanism has also been proven to be truthful [27]: if the user $i$ bids the true value $b_{ij} = v_{ij}$,

---
[3]Each bid $B_i$ is a function $A \to \mathbb{R}$.

**Mechanism 1 Shapley Value Mechanism:** Computes the users serviced by an optimization $j$, and their cost-share $p_{ij}$.

**Input:** Optimization cost $C_j$; bids $b_{1j}, \ldots, b_{mj}$.
**Output:** Serviced users $S_j$; cost shares $p_{1j}, \ldots, p_{mj}$
  $S_j \leftarrow \{1, \ldots, m\}$ /* The set of serviced users */
  **repeat**
    $p \leftarrow \frac{C_j}{|S_j|}$ /* Divide cost evenly */
    $S_j \leftarrow \{i \mid i \in S_j, p \leq b_{ij}\}$ /* Users still willing to pay */
  **until** $S_j$ remains unchanged, or $S_j = \emptyset$
  $p_{ij} \leftarrow p$ if $i \in S_j$ /* Serviced users pay the same amount */
  $p_{ij} \leftarrow 0$ if $i \notin S_j$. /* Non-serviced users do not pay */
  **return** $(S_j, (p_{ij})_{i=1,m})$

her utility (which is $v_{ij} - p_{ij}$ if $i \in S_j$, and 0 otherwise) is no smaller than her utility under any other bid. Indeed, if she underbids, *i.e.*, $b_{ij} < v_{ij}$; two cases are possible. If $b_{ij} < C_j/|S_j|$, Mechanism 1 removes her from $S_j$ and finds a smaller set of serviced users $S_j$ that excludes her: thus, her utility drops to 0. Else she continues to belong to $S_j$, so her payment $p_{ij}$ and her utility remain unchanged. Hence, she cannot increase her utility by underbidding. The reader may check that overbidding can not improve her utility either.

### 4.2 Add$^{\text{Off}}$ Mechanism

We now propose our first mechanism for cloud optimization, under the simplest setting, when the optimizations are done offline and are additive; we remove these restrictions in the next sections. Our mechanism, called Add$^{\text{Off}}$, iterates over $J$ and runs the Shapley Value Mechanism for each optimization. It adds to $a$, the grant pairs for all serviced users, and it implements the optimization $j$ when the set $S_j$ is not empty. Each user pays the sum of all per-optimization payments. Since Add$^{\text{Off}}$ runs the Shapley Value Mechanism, independently, for each optimization, it follows directly that it remains truthful and cost-recovering, as the latter.

Even though no mechanism can be truthful, cost-recovering and efficient simultaneously, the Shapley Value mechanism has the important property of *minimizing utility lost* due to the cost-recovery constraint [27]. We show, in Section 7, how this leads to high utilities even in the face of selfish users compared to existing pricing techniques.

## 5. A MECHANISM FOR DYNAMIC COLLABORATIONS

The simple offline mechanism in the previous section is insufficient for optimizations in the cloud, because cloud users change over time. In this section, we develop a new *online mechanism* for pricing cloud optimizations, where users may join and leave the system at any time. In general, a truthful offline mechanism may no longer be truthful in an online setting [31, p. 412]; similarly, applying an offline cost-recovering mechanism to an online setting may render it non cost-recovering. Our new mechanism is specifically designed for an online setting, and we prove that it is both truthful and cost-recovering. We continue to restrict our discussion to additive optimizations (we drop this assumption in the next section), and therefore, without loss of generality, we discuss the mechanism assuming a single optimization $j$.

An optimization's cost has two components: an initial implementation cost (*e.g.*, building an index) and a maintenance cost (*i.e.*, cost of index storage and index maintenance). To avoid oscillations where users can afford the initial implementation cost but not its maintenance cost, we propose an approach where the cloud computes a single, fixed cost $C_j$, for each optimization $j$. This cost captures both the initial implementation cost and the maintenance cost for some extended period of time $T$ (*e.g.*, a month). Users may join and leave at anytime during $T$. However, at the end of this time-period, the optimization's cost is recomputed and all interested users must purchase it again.

## 5.1 Add$^{\text{On}}$ Mechanism

We first explain how we model the time $T$. We divide $T$ into time-slots numbered $1 \ldots z$ where a slot is the smallest time interval for which a user can buy the service. If $T$ is a month, slots could correspond to hours, days or weeks. The value for user $i$ is a tuple $\theta_{ij} = (s_i, e_i, v_{ij})$. Here, $s_i$ is the slot when she enters the system (*e.g.*, by opening an account) and $e_i$ is the slot when she leaves the system. $v_{ij}(t)$ is the function over the slots $1 \ldots z$ such that: at each slot $t \in [s_i, e_i]$, if user $i$ gets access to the optimization $j$, she obtains the value $v_{ij}(t)$; else she obtains a value of 0. We assume that if $t < s_i$ or $t > e_i$, $v_{ij}(t) = 0$. $v_{ij}(t)$ can be an arbitrary non-negative function and may be such that user $i$ only uses the optimization for a subset of the slots in $[s_i, e_i]$.

Users bid for the optimization $j$, by declaring their values as $\theta_{ij} = (s_i, e_i, b_{ij})$, where $b_{ij}(t)$ is a function of time over the interval $t \in [s_i, e_i]$. The cloud collects the bids at each slot $t \in [1, z]$: a bid cannot be retroactive ($s_i < t$), but users are allowed to revise their future bids ($b_{ij}(t')$, $t' \geq t$) upwards[4]. For example, at time $t = 1$, let user 1 bid $(1, 3, [10, 10, 10])$, meaning $b_{1j}(1) = b_{1j}(2) = b_{1j}(3) = 10$; at time $t = 2$ she may revise her bids as $b_{1j}(2) = 20, b_{1j}(3) = 10$. For each time-slot $t$, the cloud needs to determine the set of serviced users $S_j(t)$, based on the current bids. When a user $i$ leaves the system at time $e_i$, she has to pay a certain amount $p_{ij}$.

EXAMPLE 2. *Consider an optimization $j$ with cost $C_j = 100$, and two users with values: $\theta_{1j} = (1, 1, [101])$, $\theta_{2j} = (1, 2, [26, 26])$. Thus, user 1 obtains a value of 101 at $t = 1$ if she can access the optimization; user 2 obtains a value 26 at each of the times $t = 1, 2$, if she can access the optimization. Consider the following naïve adaptation of the Shapley Value Mechanism to a dynamic setting. Run the mechanism at each time-slot, until it decides to implement the optimization: at that point the cloud has recovered the cost, and will continue to offer the optimization for free to new users. In our example, the optimization will be performed at $t = 1$, each user will pay 50, and $52 - 50 = 2$ will be user 2's utility. The problem is that the mechanism is not truthful: user 2 may cheat by bidding $(2, 2, [26])$. That is, if user 2 hides her value during the first slot, user 1 would pay the entire cost of the optimization, at $t = 1$, and user 2 would get a free ride at $t = 2$, obtaining a higher utility of $26 - 0 = 26$.*

Our mechanism addresses the challenge outlined in the above example. Mechanism 2 shows the detailed pseudo-code. Intuitively, it works as follows: First, it runs the Shapley-Value Mechanism at each slot $t$ using the *residual bid* $\sum_{\tau \geq t} b_{ij}(\tau)$ for each user $i$ (line 7). The residual bid captures the remaining value that each user would achieve if the optimization were implemented at the current slot $t$. This process repeats until the mechanism reaches a slot with a high enough value in the residual bids to implement

---
[4] As a consequence, $e_i$ can only increase.

---

**Mechanism 2 Add$^{\text{On}}$ Mechanism:** Cost-sharing mechanism for *additive* optimizations, for *multiple* slots.

**Input:** Optimization $j$; cost $C_j$; bids $(s_i, e_i, b_{ij})_{i=1,m}$.
**Output:** Serviced users $(S_j(t))_{t=1,z}$; payments $(p_{ij})_{i=1,m}$
1: $CS_j(0) \leftarrow \emptyset$ $\quad p_{ij} \leftarrow 0, \forall i = 1, m$
2: **for** each time slot $t = 1, z$ **do**
3: $\quad$ **for** each user $i = 1, m$ **do**
4: $\quad\quad$ **if** $i \in CS_j(t-1)$ **then**
5: $\quad\quad\quad$ $b'_{ij} \leftarrow \infty$ /* Force user $i$ to be serviced */
6: $\quad\quad$ **else if** $t \geq s_i$ **then**
7: $\quad\quad\quad$ $b'_{ij} \leftarrow \sum_{\tau \geq t} b_{ij}(\tau)$ /* Residual value at time $t$ */
8: $\quad\quad$ **else**
9: $\quad\quad\quad$ $b'_{ij} \leftarrow 0$ /* Prune users not yet seen */
10: $\quad\quad$ **end if**
11: $\quad$ **end for**
12: $\quad$ /* Update the set of serviced users */
13: $\quad$ $(CS_j(t), (p'_{ij})_{i=1,m}) \leftarrow$ Shapley-Mech$(C_j, (b'_{ij})_{i=1,m})$
14: $\quad$ $S_j(t) \leftarrow \{i \mid i \in CS_j(t), t \leq e_i\}$ /* Service active users */
15: $\quad$ **for** $i = 1, m$ **do**
16: $\quad\quad$ **if** $e_i = t$ **then**
17: $\quad\quad\quad$ $p_{ij} \leftarrow p'_{ij}$ /* User $i$ pays when her bid expires */
18: $\quad\quad$ **end if**
19: $\quad$ **end for**
20: **end for**
21: **return** $((S_j(t))_{t=1,z}, (p_{ij})_{i=1,m})$.

---

the optimization. At that time, the optimization is implemented, the users who could afford it get access to it, and an *initial* cost-share is computed. In subsequent time-slots, all previously serviced users continue to be serviced. If a new user arrives, the system has two options: allow her to pay the previously computed cost-share and access the optimization or *recompute a lower cost-share* given the extra contribution of the new user. We choose the latter approach since it minimizes the cost-share and maximizes the number of users who get the service. As a result, *the per-user cost-share decreases as new users join the system and contribute to the optimization cost.* Users actually pay for the optimization only when they leave the system at time $e_i$. At that time, they pay the lowest cost-share computed so far. Notice that, when a user $i$ pays and leaves, the cost-share does not increase for the remaining users since $i$ paid her share of the optimization cost.

More formally, the Add$^{\text{On}}$ Mechanism computes for each time-slot $t \in [1, z]$ the set of serviced users $S_j(t)$ (line 14), and computes the payment $p_{ij}$ (lines 15-19) for each user $i$ leaving at time $t$, using the Shapley-Value mechanism. Denote the cumulative set of serviced users as $CS_j(t) = \bigcup_{\tau \leq t} S_j(\tau)$. The key modification to the Shapley-Value mechanism is to have it operate on $CS_j(t)$ rather than $S_j(t)$ (line 13). This ensures that *all* users who have used or will use the optimization contribute equally to pay for the cost. Once a user is serviced at some time $\tau$, $i \in S_j(\tau)$, all her future bid are assumed to be $\infty$ (line 5): this ensures that the Shapley-Value Mechanism will always include $i$ in $CS_j(t)$. The users actually serviced, $S_j(t)$, are the active users in $CS_j(t)$ (line 14).

EXAMPLE 3. *Let the cost of the optimization be $C_j = 100$ with four users bidding $(1, 1, [101]), (1, 3, [16, 16, 16]), (2, 2, [26]), (2, 2, [26])$. Then $CS_j(1) = \{1\}$, $CS_j(2) = \{1, 2, 3, 4\}$, $CS_j(3) = \{1, 2, 3, 4\}$. Note that user 2 is not included in $CS_j(1)$ because her bid 48 is below $C_j/2$. At time $t = 2$ her remaining total value* is only 32: however, since now there are four users, each users' share is $C_j/4$ and therefore all users are included in $CS_j(2)$, and in $CS_j(3)$. Users 1,2,3,4 leave at times $t = 1$, $t = 3$, $t = 2$, $t = 2$ respectively, so they pay $100, 25, 25, 25$.

## 5.2 Properties

We prove that Add$^{On}$ has three important properties: (1) it is resilient to bids with both untruthful values and untruthful times, (2) it is cost-recovering, and (3) although users can increase their own utilities by using multiple identities, they can not decrease the utility of other users.

*Truthful.* The definition of a truthful mechanism in the dynamic setting is more subtle than in the static setting. In a static scenario, the mechanism is called truthful if for *any set of bids*, user $i$ cannot obtain more utility by bidding $b_{ij} \neq v_{ij}$ than by bidding her true value $b_{ij} = v_{ij}$. In the dynamic case, user utilities depend not only on the other bids received until now, but also on what will happen in the future. We assume the *model-free* [31] framework to define truthfulness in the dynamic case: it assumes that bidders have no knowledge of the future agents and their preferences. At each time $t$, every agent assumes their worst utility over all future bids, and they bid to maximize this worst utility [31].

EXAMPLE 4. *Consider Example 3. User 2 bids $(1, 3, [16, 16, 16])$, thus she could obtain a value 16 at each of the three time-slots $t = 1, 2, 3$; but she is serviced only at time-slots $t = 2, 3$, hence her value is $16 + 16 = 32$. She pays 25, thus her utility is $32 - 25 = 7$. Suppose that she cheats, by overbidding $(1, 3, [17, 17, 17])$. Now she is serviced at all three time-slots, but still pays only 25 (because when she leaves there are four users in $CS_j$). Thus, for the particular bids in Example 3, user 2 could improve her utility by cheating. In a model-free framework, however, users do not know the future, and they must assume the worst case scenario. In our example, the worst case utility for user 2 at $t = 1$ (when she places her bid) corresponds to the case when no new bids arrive in the future: in this case, if she overbids $\geq 50$, she ends up paying 50, and her utility is $48 - 50 = -2$. If she underbids, her worst case utility is still 0. By cheating at $t = 1$, user 3 cannot increase her worst-case utility.*

With the *model-free* notion of truthfulness, a dynamic mechanism is called truthful if, for each user, revealing her true preferences maximizes the minimum utility that she can receive, over all possible bids by future users. This definition of truthfulness reduces to the classic definition of truthfulness for the static case (*i.e.*, with a single time slot).

PROPOSITION 1. *Add$^{On}$ Mechanism is truthful.*

PROOF. (Sketch) Consider a user $i$ bidding at time $t$, *i.e.*, her bid is $(s_i, e_i, b_{ij})$ and $t \leq s_i$ (bids cannot be placed for the past). We claim that her minimum utility over all future users' preferences (at times $t + 1, t + 2, \ldots$) is when no new bids arrive in the future. Indeed, any new bids in the future can only decrease the payment due by user $i$ (by increasing the set $S_j(e_i)$, hence decreasing her payment $p_{ij} = C_j/|S_j(e_i)|$), and can only increase her value at every future time slot $t' \leq s_i$, by including $i$ in a set $S_j(t')$ where it was previously not included. Thus, the minimum utility for user $i$ is when no new bids arrive after time $t$. But in that case, Add$^{On}$ degenerates to one round of the Shapley-Value Mechanism, run at time $t$, which is proven to be truthful. □

*Cost-recovering.* Intuitively, Add$^{On}$ recovers all costs because it always applies the Shapley-Value Mechanism to the game given by all bids known at the present time. Due to the lack of space, we defer the proof to our technical report [41].

*Multiple Identities.* A user could create multiple identities and place a separate bid for each identity. If at least one identity gets access to the optimization, she obtains her full value (by running her queries under that identity). However, she has to pay on behalf of all identities. It turns out that a user can increase her utility this way: by creating more identities, she could help more users to be serviced and thus decrease her total payment. For example, consider an optimization that costs $C_j = 101$ and a user Alice whose value is $(1, 1, [101])$. Suppose there are 99 other users whose values are $(1, 1, [1])$. Of the 100 users, only Alice is serviced, because even if all the other 99 users were serviced, each would be paying $101/100 = 1.01$, which would exceed their value of 1. However, if Alice creates two identities, each bidding (say) $(1, 1, [101])$, Add$^{On}$ would see 101 users and would serve all of them with each of the 99 users paying $101/101 = 1$, while Alice would pay 2, once for each identity. Thus, her utility would increase from $101 - 101 = 0$ to $101 - 2 = 99$. Add$^{On}$ does not prevent such ways of gaming the system, because they are indistinguishable from collaborations. For example, instead of cheating, Alice could ask Bob (whose value is at least 1) to participate in the game, then reimburse Bob for his payment: this is indistinguishable from creating a fake identity. On the other hand, this is not undesirable: through her action, she caused more users to be serviced, while agreeing to pay a bit more than the other users' shares. We can prove that this holds in general.

PROPOSITION 2. *Suppose a user $i$ can increase her utility under Add$^{Off}$ or Add$^{On}$ by creating multiple identities $i_1, i_2, \ldots$. Then no other users' utility decreases.*

PROOF. (Sketch) Consider two games, one with user $i$ with a single account and one with user $i$ creating $k$ identities $i_1, \ldots, i_k$ and associated bids. Her utility can increase by creating dummy identities only if the total payment by the dummies is less than the total payment without the dummies. Let user $i$'s payment with no dummies be $p_i$ and the total payment of her dummies be $p'_i$. Since creating dummies increases $i$'s utility $p'_i < p_i$, and the payment per dummy (which would be the payment per user as well with the dummy accounts) is $p'_i/k < p'_i < p_i$. Thus, for all users served in the game with no dummies are surely served with dummies too since the payment per user is lower than without the dummies. Hence the utility of no user decreases. □

## 6. MECHANISMS FOR SUBSTITUTABLE OPTIMIZATIONS

In this section, we relax the requirement that optimizations be independent. Indeed, when multiple optimizations (*e.g.*, indexes or materialized views) exist, the value to the user from a set of optimizations can be a complex combination of the individual optimization values. In this section, we consider the case of substitutable optimizations. Formally, each user defines a set of substitutable optimizations $J_i \subseteq J$ such that $\forall j, k \in J_i : v_{ij} = v_{ik} = v_i > 0$. Additionally, given an outcome $a$, $V_i(a) = v_i$ if $\exists j \in J_i : (i, j) \in a$ and $V_i(a) = 0$ otherwise. In comparison to the substitutable valuation, the valuation function that we previously used was

**Mechanism 3 Subst^Off Mechanism:** Cost-sharing mechanism for *substitutable* optimizations for a *single* slot.

---

**Input:** Opts. $J$; costs $(C_j)_{j=1,n}$; bids $(b_{ij})_{i=1,m;j=1,n}$
**Output:** Alternative $a \in A$; cost shares $(p_{ij})_{i=1,m;j=1,n}$
  $a \leftarrow \emptyset \quad p_{ij} \leftarrow 0, \forall i = 1, m \quad \forall j = 1, n$
  **loop**
    **for** each optimization $j$ in $J$ **do**
      /* Compute serviced users, discard payments */
      $(S_j, (p'_{ij})_{i=1,m}) \leftarrow$ Shapley-Mech$(C_j, (b_{ij})_{i=1,m})$
    **end for**
    /* Find the smallest cost-share optimization */
    $J^f \leftarrow \{j \in J \mid S_j \neq \emptyset\}$ /* Set of feasible opts */
    **if** $J^f \neq \emptyset$ **then**
      $j_{min} \leftarrow \arg\min_{j \in J^f}(C_j/|S_j|)$
      $a \leftarrow a \cup \{j_{min}\}$ /* Perform optimization $j_{min}$ */
      **for** each user $i \in S_{j_{min}}$ **do**
        $a \leftarrow a \cup \{(i, j_{min})\}$
        $p_{ij_{min}} \leftarrow C_{j_{min}}/|S_{j_{min}}|$
        $b_{ij} \leftarrow 0 \quad \forall j \in J$ /* Remove $i$ from future loops */
      **end for**
      $C_{j_{min}} \leftarrow \infty$ /* Remove $j_{min}$ from future loops */
    **else**
      **return** $(a, (p_{ij})_{i=1,m;j=1,n})$
    **end if**
  **end loop**

---

**Mechanism 4 Subst^On Mechanism:** Cost-sharing mechanism for *substitutable* optimizations, for *multiple* slots.

---

**Input:** Opts $J$; costs $(C_j)_{j=1,n}$; bids $(s_i, e_i, (b_{ij})_{j=1,n})_{i=1,m}$.
**Output:** Serviced users $(S_j(t))_{t=1,z}$; payments $(p_{ij})_{i=1,m}$
  $a \leftarrow \emptyset \quad p_{ij} \leftarrow 0, \forall i = 1, m$
  **for** each time slot $t = 1, z$ **do**
    **for** each user $i = 1, m$ **do**
      **if** $\exists j \in J. (i, j) \in a$ **then**
        $b'_{ij} \leftarrow \infty$ /* Force user $i$ to be serviced */
        $b'_{ij'} \leftarrow 0 \quad \forall j' \in J, j' \neq j$ /* Force $i$ to only use $j$ */
      **else if** $t \geq s_i$ **then**
        $b'_{ij} \leftarrow \sum_{\tau \geq t} b_{ij}(\tau)$ /* Remaining value know at t */
      **else**
        $b'_{ij} \leftarrow 0$ /* Prune users not yet seen */
      **end if**
    **end for**
    /* Update the set of serviced users */
    $(a, p'_{ij}) \leftarrow$ Subst$^{Off}(J, (C_j)_{j=1,n}, (b'_{ij})_{i=1,m;j=1,n})$
    $S_j(t) \leftarrow \{i \mid \exists j. (i, j) \in a, t \leq e_i\}$
    **for** $i = 1, m$ **do**
      **if** $e_i = t$ **then**
        $p_{ij} \leftarrow p'_{ij}$ /* User $i$ pays when her bid expires */
      **end if**
    **end for**
  **end for**
  **return** $((S_j(t))_{j=1,n;t=1,z}, (p_{ij})_{i=1:m,j=1:n})$

---

the *sum*: $V_i(a) = \sum_{(i,j) \in a} v_{ij}$. With substitutable valuations, a user bid takes the form $\theta_i = (J_i, v_i)$, where $J_i$ is the set of substitutable optimizations and $v_i$ is the user value if she is granted access to at least one optimization in $J_i$.

Substitutable optimizations capture the case where implementing any optimization from a set (*e.g.*, indexes, materialized views, or replication) can speed-up a workload by a similar amount and the user does not have any preference as to which optimization is responsible for the speed-up. However, she gets no added value from multiple optimizations being implemented at the same time either because they may be redundant (*e.g.*, a materialized view may remove the need for a specific index) or because she is indifferent to further performance gains.

### 6.1 Subst^Off Mechanism

We first consider the Subst^Off Mechanism for static games where all users use the system for the same time period.

EXAMPLE 5. *Consider three optimizations with costs $C_1 = 60$, $C_2 = 180$, and $C_3 = 100$. The bid $(\{1, 2\}, 100)$ indicates that the user values the access to either optimization 1 or 2 at 100. Other example bids include $(\{3\}, 101)$, $(\{1, 2, 3\}, 60)$, and $(\{2\}, 70)$, for users $\{2, 3, 4\}$, respectively.*

The challenge with substitutable optimizations is that users may bid for partially overlapping sets of optimizations as in Example 5. They also have a new way of cheating. In addition to lying about their value $v_i$ and emulating multiple users, they may lie about the optimizations they want by either bidding for ones they do not want or by not bidding for the ones they do want. Our mechanisms are truthful under the *model-free* notion and are also resistant to cheating with dummy users under the practical assumption that no user knows other users' bids.

Subst^Off Mechanism (Mechanism 3) works in a sequence of phases. In the first phase, it runs the Shapley Value mechanism for each optimization $j$ (along with the users who bid for $j$) independently and selects the optimization $j_{min}$ with the lowest cost-share. Users who want $j_{min}$ and can pay its cost-share get access to it. The mechanism then recursively applies the algorithm to the remaining users and optimizations in subsequent phases.

EXAMPLE 6. *Consider example 5. Subst^Off first identifies optimization 1 as having the lowest cost-share with $S_1 = \{1, 3\}$ and cost-share $\frac{60}{2} = 30$, and thus implements optimization 1 and services users 1 and 3. Next, Subst^Off considers the remaining users $\{2, 4\}$ and the remaining optimizations $\{2, 3\}$. For these optimizations, $S_2 = \emptyset$ while $S_3 = \{2\}$. Optimization 3 is thus implemented and user 2 is given access to it. User 4 gets access to no optimization.*

Due to space constraints we defer the proof that Subst^Off is cost-recovering and truthful to our technical report [41]. Example 7 provides an intuition for its truthfulness.

EXAMPLE 7. *Consider example 6. If, to cheat, user 3 bids any value in the range $[30, \infty)$, the outcome and her utility would not change. If she bids below 30, however, she would not be serviced by optimization 1 as her bid would be below the cost-share. She would not get serviced by any other optimization either, because their cost-shares are higher than that of optimization 1, which has the lowest cost-share. Her utility would be $(0 < 30)$. Finally, if she, being untruthful, does not bid for optimization 1, even though it benefits her, and bids $(\{2, 3\}, 60)$, then both optimization 1 and 2 would tie for the lowest cost-share at 60. Assuming that Subst^Off makes a random choice and implements optimization 2, then she would get access to optimization 2 and would pay the cost-share of 60, achieving a strictly lower utility of 0.*

### 6.2 Subst^On Mechanism

We now consider substitutable optimizations, but in a dynamic setting where users can join and leave the system in any time-slot. Given substitutable optimizations $J_i$, user $i$

568

bids $\omega_i = (s_i, e_i, b_i, J_i)$, with $[s_i, e_i]$ as the requested interval of service and $b_i(t)$ is the value she gets at time $t$.

Subst$^{On}$ Mechanism, shown in Mechanism 4, works by running Subst$^{Off}$ at each time-slot $t$ with the residual value of all the users seen. The first time a user $i$ is granted access to optimization $j$ her bid for $j$ is updated to $\infty$ (so that she is always in the feasible set of $j$), while her bids for the other optimizations are updated to 0 (so that she remains serviced only by optimization $j$).

EXAMPLE 8. *Consider three optimizations $\{1, 2, 3\}$ with costs $C_1 = 60, C_2 = 100, C_3 = 50$. User 1 bids $(1, 2, 100, \{1, 2\})$, which is interpreted as follows: she values any optimization in $\{1, 2\}$ at 100 for the time-slots $[1, 2]$. User 2 bids $(2, 3, 100, \{1, 2, 3\})$ and user 3 bids $(3, 3, 100, \{3\})$. At $t = 1$, Subst$^{On}$ runs Subst$^{Off}$ with user 1 (the only user at that time) and ends up implementing optimization 1, with a payment of 60. Then, Subst$^{On}$ updates user 1's bid to optimization $\{1\}$ valued at $\infty$. At time $t = 2$, Subst$^{On}$ runs Subst$^{Off}$ with users $\{1, 2\}$ and ends up granting user 2 access to optimization 1 with the new payments for both users being $60/2 = 30$. User 1 leaves after paying 30, while user 2's bids are updated to optimization $\{1\}$ valued at $\infty$. At time $t = 3$, Subst$^{On}$ again executes Subst$^{Off}$ with all three users (although user 1 left, she is included while invoking Subst$^{Off}$, to compute the proper cost-share for user 2), and ends up implementing optimization 3, but only for user 3, at a payment of 50. User 2 is not serviced optimization 3 since she is already using optimization 1 and Subst$^{On}$ does not allow her to switch to a new optimization. The system ends with user 2 paying 30 and user 3 paying 50. The inability to switch is crucial for truthfulness: otherwise, a new user, say user 4, who prefers optimization $\{1, 3\}$, arriving at time $t = 3$, might only bid for optimization 3 hoping that user 2 would switch to optimization 3. If user 2 could switch, each would pay $50/3 = 16.7$, while without the switch, user 2 pays $60/2 = 30$ (as before) and users $\{3, 4\}$ pay $50/2 = 25$.*

Due to space constraints we defer the proof that Subst$^{On}$ is truthful and cost-recovering to our technical report [41].

*Multiple Identities.* The dummy users can, in theory, increase their utility at the expense of other users, for substitutable optimizations, though this is hard to do in practice. We illustrate this for Subst$^{Off}$, but the conclusions also apply to Subst$^{On}$. Consider users $\{1, 2, 3\}$ with single-slot bids $(\{1\}, 5)$, $(\{1, 2\}, 2.51)$, and $(\{2\}, 7)$ for optimizations $\{1, 2\}$ with costs $C_1 = 6$ and $C_2 = 5$. With no dummy users, optimization 2 is implemented with a payment of 2.5 and utilities of 0.01 for user 2 and 4.5 for user 3. If user 1 creates two identities $1'$ and $1''$ that make a bid of 2.5 each for optimization 1, then both optimizations are implemented with optimization 1 serving $\{1', 1'', 2\}$ with utilities of 1, 0.51, and 2 for users 1, 2, and 3 respectively. Note that user 3's utility has reduced. However, to cheat, user 1 needed to know the number of other users and their bids, which is not publicly known in practice. She may try guessing, but in the worst case, her guess can lead to a reduction in her utility [41]. Thus, being truthful is the optimal strategy when the user does not know the other bids.

## 7. EVALUATION

Our mechanisms guarantee truthfulness and cost-recovery, but they do not optimize for total utility. In this section, we empirically evaluate the total utility that our solutions provide. We focus on the two online mechanisms (*i.e.*, Add$^{On}$ Mechanism and Subst$^{On}$ Mechanism) and compare them to the state-of-the-art regret-based approach (Section 7.1) [16, 22]. The experiments consist of both the motivating use-case (Section 2) and simulated scenarios (Sections 7.3 through 7.6).

### 7.1 Regret-based Amortization

Kantere, Dash, *et al.* [16, 22] proposed a regret-based approach (called Regret, henceforth) to select optimizations. They developed a detailed economy of the cloud and considered detailed query plans for computing regret. In this paper, we abstract away and evaluate the performance of the core regret-based approach without the surrounding economy or plan details. We briefly describe the algorithm.

The regret for an optimization $j$ at time $t$, termed $R_j(t)$, is defined as the total *value* that would have been realized, over all users, until time $t$ (and excluding time $t$), had $j$ been implemented at $t = 0$. Formally, $R_j(t) = \sum_{\tau < t} \sum_{i \in I} v_{ij}(\tau)$, where $I$ is the set of all users and $v_{ij}$ is user $i$'s valuation for optimization $j$. The policy we adopt is the greedy approach [31] where the optimization is implemented at that time-slot $t$ when $c_j \leq R_j(t)$. For substitutable optimizations, once an optimization $j$ is implemented for a user $i$, she stops benefiting from the other optimizations $J \setminus \{j\}$ and does not contribute to their regret.

We now explain how Regret sets prices. For ease of explanation we assume a single optimization $j$ that Regret implements at time $t_r$. Users in subsequent time-slots can get access to it only after paying a price $p_j$. Regret chooses $p_j$ to be the minimum payment such that the total payment from future users equals $c_j$. If no price $p_j$ can recover the cost, it picks a price that minimizes the cloud's loss. Note that Regret uses the residual value in the game assuming *perfect knowledge* of future users' values. If $I_j(p, t_r) = |\{i | \sum_{t > t_r} v_{ij}(t) \geq p\}|$ is the number of future users who would pay $p$ for optimization $j$, then the cloud-loss would be $L_j(p, t_r) = (c_j - pI_j(p, t_r))$. The payment $p_j$ minimizes this loss, *i.e.*, $p_j = \arg\min_p \max\{L_j(p, t_r), 0\}$. (Choose the smallest $p_j$, in case of ties, so that user utilities are maximized.) Thus, our price point is the optimal choice to minimize the cloud-loss: it gives an *upper bound* on how well Regret would work in practice. The total social utility (*a.k.a.* total utility) for Regret is defined the same way as for the mechanisms (Section 3): the total value realized by the users for the slots they are serviced minus the implemented optimizations' costs. The *cloud balance* is the costs of the optimizations minus the total payments by the users. A negative balance means that the cloud incurs a loss.

Our approach thus computes regret the same way as Kantere, Dash, *et al.* [22, 16] except that, in their approach, users assign values to individual queries. Our approach aggregates this information and assigns values to workloads spanning larger periods of time.

### 7.2 Evaluation on the Motivating Use-Case

The workload from the motivating use-case in Section 2 traces the evolution of halos over 27 snapshots of a universe simulation. Each astronomer starts with a subset of halos, $\gamma$, in the final snapshot at $t_{27}$ and, for each halo $g \in \gamma$, she (a) computes the halos in each previous snapshot contributing the most *particles* to $g$, and (b) recursively computes a chain of halos $(h_1^g, \ldots, h_{26}^g, g)$ such that $h_t^g$ contributes the most *mass* to the halo $h_{t+1}^g$ in the next snapshot. Our optimizations materialize the following relation for each snapshot: (particleID, haloID) to speed-up the queries.

We experiment with six users with differing workloads: two workloads (in use by the astronomers) trace the evolution of halos $\gamma_1$ and $\gamma_2$, respectively, using all 27 snapshots. Based on the astronomers' feedback, we define two new users for each of $\gamma_1$ and $\gamma_2$: one user uses every $2^{nd}$ snapshot while the other uses every $4^{th}$ snapshot. This simulates faster, exploratory studies of the data. In our experiments, we measure the total utility (Sec. 3) for both Add$^{On}$ and Regret.

We take each optimization's cost to be the dollar amount of storing the materialized view on a yearly subscription of the Amazon EC2 High-Memory Extra Large Instance [5]. This yields an average cost of $2.31 per optimization.[5]

We take the money saved, by completing queries earlier, to be the value of an optimization (Amazon also charges for each hour of use in addition to the subscription fee). For the six users, the run-time of their workload without any optimizations is 81, 36, 16, 83, 44 and 17 mins. Materializing the view on the snapshot 27 saves 44, 18, 8, 39, 23, and 9 min which corresponds to monetary savings of 18, 7, 3, 16, 9, and 4 cents for one execution of the workloads. The other optimizations reduce run-time by 2.5 min each for a saving of 1 cent. Since the optimizations affect different queries in the workload, we take them to be additive.

We consider a year-long time-period where each user uses the service in multiples of a quarter (3 months). We explore *all* the $10^6$ ways that the group can bid for slots. For each alternative, we then vary the total number of executions of each user's workload, and we compute the total utility achieved by each approach. Figure 1 shows the average and the standard deviation of the utilities across the $10^6$ alternatives as we change usage intensity from low (1 workload execution/quarter) to medium (1 workload execution/day).

Compared to the baseline cost, taken to be the total cost of executing the workloads without optimizations, Add$^{On}$ and Regret yield total utilities of 28%-47% and 16%-40% of the base line cost, respectively. Since Add$^{On}$ ensures that users will pay the entire cost, the total utility is exactly the amount of money saved by the group; while for Regret, the total money users save is the sum of the total utility and the unpaid fraction of the cost, *i.e.* the cloud balance. We add this balance to the utility since the total utility includes the utility of both the users and the cloud.[6] Thus, both approaches significantly reduce the cost of using the cloud.

Comparing Add$^{On}$ to Regret, we find that Add$^{On}$ yields a total utility that is 18%-118% higher than Regret, at 90 and 40 executions per user, respectively. Further, while the cloud never makes a loss with Add$^{On}$, loss by Regret can be up to a substantial 92% of Regret's utility (at 40 executions). As noted before, our outcomes for Regret are an upper bound and with more realistic bids Regret is likely to do even worse.

---
[5]We could have used a different instance. We chose this one as it was the most similar to our local machine, on which we obtained the storage space and query run-time values.

[6]In the case of a scientific collaboration, we can also assume that one of the researchers pays a public cloud to implement the optimization. She then asks the other researchers to pay her back. That researcher is then the one who incurs the loss. In this case, the total social utility would be the amount saved by the entire group of researchers.

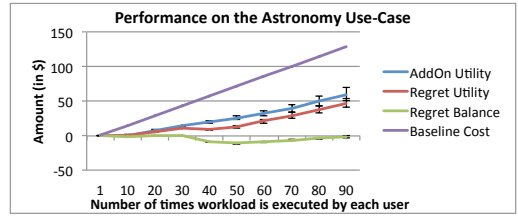

Figure 1: Operating *expenses* without optimization and total utility (equal to total money saved) by Add$^{On}$ and Regret for the astronomy workload on an Amazon EC2 subscription, as workloads are executed more frequently.

In practice, users would execute their workloads multiple times and datasets are likely to be larger. For example, the upcoming NCSA/IBM Blue Waters system [28] can generate 10 TB to 200 TB per snapshot (as opposed to 4.8 GB per snapshot for our experiments). With a 3 to 5 orders of magnitude increase in data size, building optimizations and executing workloads would be correspondingly costlier, and sharing optimizations would lead to proportionately larger savings in the order of tens of thousands of dollars.

## 7.3 Collaboration Size

In the remaining sections, we use a variety of simulated configurations to explore how our mechanisms and the Regret approach compare in different settings. In all cases, we measure the total utility.

The first key parameter affecting utility is the cost of optimizations as a proportion of the user values. This ratio affects the number of users that are necessary to cover the optimizations' cost. In all simulations, we change this proportion by varying the per-optimization cost along the x-axis while keeping the average user values constant. In this section, we measure the utility of both approaches when the total number of users available to cover the optimizations' cost is either small (small collaborations) or large (large collaborations). For both approaches, users in larger collaborations can buy costlier optimizations to get higher utilities. We experiment with a small group of 6 users and a large one with 24 users. We let users pick *one* service slot, uniformly at random, from 12 slots[7]. This gives us an expected number of users/slot of 0.5 and 2, respectively.

### 7.3.1 Additive Optimizations

We first consider additive optimizations. We only consider one optimization since optimizations are independent.

For small collaborations, Figure 2(a) shows that as we move from cheap to costly optimizations, Regret provides good total utility, but then quickly leads to cloud loss, followed by negative total utility; while Add$^{On}$ never leads to cloud loss or negative utilities. Negative utilities by Regret imply that the optimization was implemented but it failed to provide enough value to justify its implementation. Restricting our attention to the costs where Regret yields a positive utility, Add$^{On}$ achieves an average total utility 1.43× higher than Regret. Further, while Regret leads to cloud loss (curve "Regret Balance" in the figure) at a cost of 0.18, even for optimizations 7× costlier, Add$^{On}$ yields substantial utility (taken to be 0.3, 10% of total user value). Regret underperforms against Add$^{On}$ for two reasons. First, for cheap optimizations that should be implemented, Regret loses user value while building up regret. Second, for costly optimizations, Regret suffers a loss and negative total utility since it implements the optimization even when the available future values is insufficient to recoup the cost.

---
[7]The number 12 was chosen since 2, 3, 4, and 6 divide it perfectly and give us a larger space of parameter values to experiment with as compared to some other number like 10 or 15. The other parameter values were chosen to be multiples of 12 for ease of understanding.

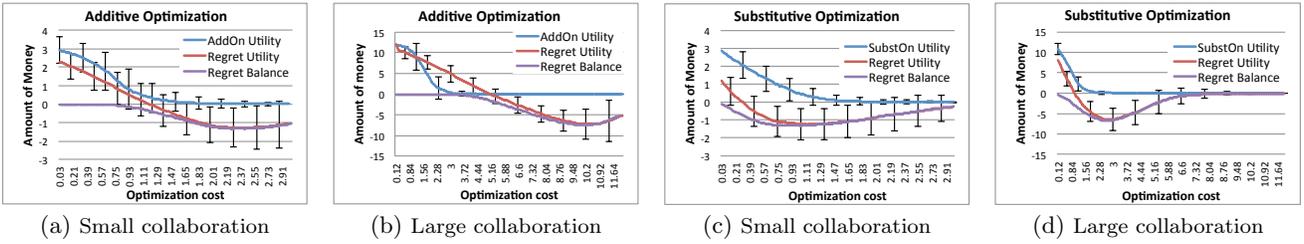

(a) Small collaboration  (b) Large collaboration  (c) Small collaboration  (d) Large collaboration

Figure 2: Total utility as a function of optimization cost for different collaboration sizes. Also showing regret balance (optimization costs minus user payments). Add$^{On}$ and Subst$^{On}$ outperform Regret for a large range of optimization costs, for both additive and substitutive optimizations, and for both low and high degrees of collaboration amongst users. Further, they never incur a loss, while Regret can incur significant loss. Detailed analysis in Section 7.3.

For larger collaborations, Figure 2(b) shows that as we move to costlier optimizations, Add$^{On}$ provides worse utility than Regret. Intuitively, Add$^{On}$ looses some opportunities to implement optimizations because it is more cautious than Regret: to avoid losses, Add$^{On}$ only implements an optimization when it is certain to recoup the costs given *current* information. The benefit of Regret, however, is limited: Regret soon starts losing money and leads to negative total utility. In fact, only in less than 10% of the range where Regret achieves a positive utility ([0, 4.92]), does it also outperform Add$^{On}$ *and* yield no loss. Over the entire range of costs in [0, 3.0] the average total utility of Add$^{On}$ is 0.87 while that of Regret is −0.63.

For large collaborations, Add$^{On}$ utilities sharply decrease after a point because when costs increase, the payment per user increases super-linearly, since Add$^{On}$ prunes out users for whom the payments are larger than the value. No users are pruned by Regret and thus it sees a linear reduction in utilities with increasing costs.

Interestingly, the range of costs for which Regret makes a loss depends on the number of users who bid. It yields a loss at a cost of 0.18 for the small group (Figure 2(a)) and 1.80 for the large one (Figure 2(b)). Thus, without knowing the future users, the cloud can not know when to avoid Regret.

#### 7.3.2 Substitutive Optimizations

To compare Subst$^{On}$ and Regret in the case of substitutive optimizations, we consider a scenario with 12 optimizations. Each user selects 3 optimizations, uniformly at random, as the set of substitutes (Section 7.6 experiments with other ratios). Unlike the additive case, the costs of the 12 optimizations are sampled uniformly from [0, 2c] so that c is the average optimization cost: this is to simulate that not all substitutes are equally expensive. Thus the x-axes of Figures 2(c) and 2(d) are the mean value of the optimizations.

Compared to the corresponding additive optimizations in Figures 2(a) and 2(b), both Subst$^{On}$ and Regret achieve lower overall utility. Indeed, with substitutes, each optimization has fewer users bidding for it and, once an optimization is implemented, the serviced users no longer pay for the other optimizations. Hence, fewer optimizations are implemented and, in the case of Regret, there are fewer users over whom the costs can be amortized. In the scenarios

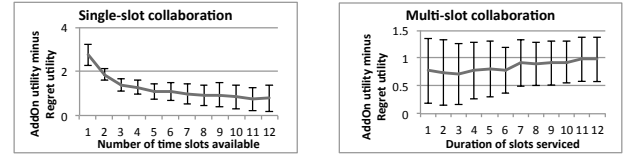

(a) More collaboration on the left. x-axis is the total number of slots. The users bid for 1 slot.

(b) Less collaboration on the left. X-axis is the # of contiguous slots that each user bids for.

Figure 3: Add$^{On}$ vs Regret performance with varying degree of collaboration. (Section 7.4)

shown, Regret yields a loss earlier than in the additive case. When averaged over those costs for which Regret yields positive utility, Subst$^{On}$ yields 1.63× and 3× more utility than Regret for group sizes of 24 and 6, respectively.

### 7.4 Overlap in Usage

The second key parameter that affects utility is how the user values are distributed across time. We study this parameter using a small group of 6 users collaborating on a single, additive optimization. We vary the degree of user overlap and its manner. First, we repeat the experiment from Figure 2(a) while decreasing the total number of slots from 12 to 1. Figure 3(a) shows that, with fewer slots to sample from and hence with increased overlap amongst users, Add$^{On}$ generates 0.77 to 2.75 more utility, on average, than Regret. Thus, Add$^{On}$ gets 25%-91% of the total user value (3.0) as additional utility over Regret. Decreasing the number of slots, increases the probability that Add$^{On}$ finds enough value in some slot to justify implementing the optimization. In contrast, regret accumulation stays unchanged.

Next, we study what happens when user values are spread across an interval rather than being concentrated in a single time-slot. The setup in Figure 3(b) is identical to the additive case with the group size of 6 in Figure 2(a) except that instead of bidding for only one slot, users bid as $(s_i, s_i + d - 1)$, where $d$ is the duration of the service and is varied on the x-axis. $s_i$ is chosen uniformly at random from 12 slots. Users divide their values, chosen uniformly at random from [0, 1), equally among all $d$ time slots in their bids. The average extra value that Add$^{On}$ generates over Regret increases from 0.77 to 0.98. Indeed, as users spread their value across multiple time-slots, Add$^{On}$ becomes more likely to find a single time-slot with sufficient value to justify implementing the optimization.

### 7.5 Arrival Skew

We now consider the small collaboration of 6 users bidding for a single optimization, where they arrive: (a) *uniformly* at random in one of 12 slots, (b) *early* following an exponential

571

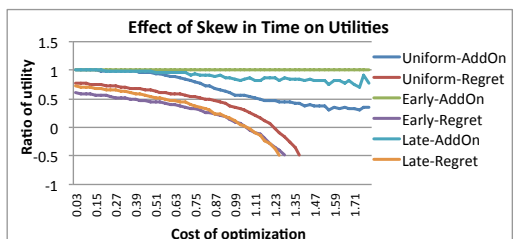

Figure 4: Add$^{On}$ improves while Regret worsens with temporal skew. Ratios taken with the utility of Add$^{On}$ with users clustered early. (Section 7.5)

distribution with mean $1.2^8$, (c) *late* following a distribution that is $12 - t$ with $t$ sampled exponentially with mean 1.2. Case (b) simulates datasets that become stale, while (c) simulates datasets that become popular over time. We look at the ratio of the utility in different settings to that of the utility of Add$^{On}$ with *early* arrivals. Figure 4 shows that total utility by Add$^{On}$ improves while that for Regret worsens with irregular arrivals. Add$^{On}$ outperforms Regret substantially as user arrival becomes non-uniform (and Regret soon starts generating negative utilities). With skew, Add$^{On}$ improves due to increased chances of finding a slot with enough value to pay for all costs. For *e.g.*, with Add$^{On}$, early arrivals can be $6.7\times$ and $1.8\times$ more efficient that uniform and late, respectively. On the other hand, Regret worsens since skew increases the chance that more regret is accumulated than required[9]. For *e.g.*, with Regret, at the cost of 0.54, late and uniform arrivals have 16% and 40% higher total utility than early arrivals, respectively. This points to an interesting property of the mechanism-design-based approach: the approach performs much better as non-uniformity increases.

### 7.6 Selectivity of Substitutes

We now vary the selectivity of the substitutes, that is defined as the ratio of the number of substitutable optimizations to the total number of optimizations. Figures 5(a) and 5(b) show the total utility for selectivities of 0.75 and 0.25, where each user chooses 3 optimizations uniformly at random from 4 and 12 optimizations, respectively. The figures show that, with more selective users, absolute utilities derived by both algorithms decrease. For *e.g.*, Regret goes from a utility of 1.10 to -0.23 while Subst$^{On}$ goes from 2.38 to 1.90 for the optimization cost of 0.36 as selectivity increases. Indeed, with more selective users, the number of users per optimization decreases and more optimizations have to be be implemented to satisfy the users. For Figures 5(a) and 5(b), Subst$^{On}$ yields an average total utility of 1.0 for optimizations that are $2.5\times$ and $12.5\times$ costlier than those at which Regret generates utilities of 1.0, respectively.

**Summary**. In summary, our mechanism-based approaches not only guarantee truthfulness and cost-recovery but also yield utility that frequently exceeds that of Regret. Our approaches work especially well in scenarios where many users derive significant value from an optimization during the same time-slot. They under-perform compared to Regret in scenarios where users value the same optimization but during non-overlapping periods.

---

[8]With mean 1.2, the maximum starting time slot of 6 users in 1000 runs was 12 as it is in case (a).

[9]Regret is computed after every time slot hence it increases in discrete values. The difference in regret and the optimization cost is wasted value and is smaller for uniform arrival.

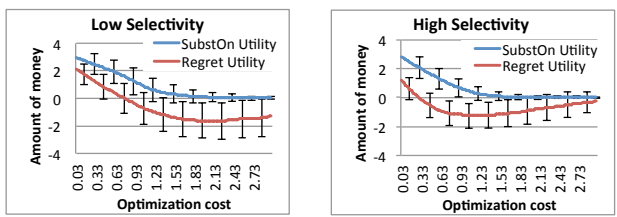

(a) Each user chooses 3 uniformly random optimizations out of 4.

(b) Each user chooses 3 uniformly random optimizations out of 12.

Figure 5: Effect of change in selectivities of substitutable optimization on total utility. (Section 7.6)

## 8. RELATED WORK

Today, cloud providers use two strategies for pricing optimizations. In the first, the cost of the optimization is included in the base service price. For *e.g.*, Amazon SimpleDB [9] automatically indexes user data and includes the corresponding overhead in the base-price computation (45 bytes of extra storage are added to each item, attribute, and attribute-value). Similarly, SimpleDB and SQL Azure [26] automatically replicate data and include that cost in the base service cost. The key limitation with this approach is that the cloud must decide up-front what optimizations are worth offering and it forces users to pay for these optimizations. In other cases, users choose desired optimizations and pay their exact cost. For example, in Amazon RDS [6] a user can choose to launch and pay-for a desired number of read-replicas to speed-up her query workload. This approach, however, works well only in the absence of collaborations.

Significant recent work studies existing cloud pricing schemes, economic models, and their implications [24, 39, 44]. In contrast we develop a new pricing mechanism.

Most closely related to our work, Dash *et al.*, developed an approach for pricing data structures (indexes, materialized views, etc.) in a DBMS cloud cache [16]. In their approach, the cloud selects the structures to build based on the notion of regret and its cost is amortized over the first $N$ queries that use it. To compute regret, the cloud relies on user supplied budget functions, that indicate their willingness to pay for various quality of service. In follow-up work Kantere *et al.* [22] tuned their approach and developed a regression-based technique to predict the extent of cost amortization. In contrast to our work, this previous approach relies on users being truthful and does not guarantee that the cost will be recovered. For example, consider a user who needs to run one, very expensive query over a private dataset. No structure will be implemented if she is truthful. Instead, she thus submits a large number of inexpensive queries over the same dataset while she expresses her willingness to pay zero for processing the extra queries, yet indicates a preference for low execution times over low costs. The regret-based approach will let her manually pick slow and cheap service for these queries. It will then compute the maximum possible regret for the missing data structure that would have enabled faster plans for these queries. When the cloud accumulates enough regret, she can run the expensive query and pay a small fraction of the total cost of the optimization.

Significant research applies economic principles to resource allocation in distributed systems [1, 12, 13, 14, 18, 34, 36, 43], collaboration promotion in peer-to-peer systems [30, 29, 42], or more recently, VM allocation in the cloud [40]. We study how to choose and price optimizations rather than allocate processing resources. The Mariposa distributed database system [38] introduced a micro-economic paradigm for optimizing distributed query evaluation and data placement. This is a problem orthogonal to ours.

We build on the Shapley Value Mechanism, which is an instance of Moulin Mechanisms [27] that have been designed for various *offline* combinatorial cost-sharing problems [32]. We design Moulin mechanisms in an online setting.

Online mechanisms [31, Ch. 16] consider games where valuations come one at a time. While there is work on characterizing truthful mechanisms to maximize social utility in dynamic games [31, Thm. 16.17], to the best of our knowledge, no work applies to cost-sharing in dynamic games.

## 9. CONCLUSIONS

We studied how a cloud data service provider should activate and price optimizations that benefit many users. We have shown how the problem can be modeled as an instance of cost-recovery mechanism design. We also showed how the Shapley Value mechanism solves the problem of pricing a single optimization in an offline setting. We then developed a series of mechanisms that enable the pricing of either additive or substitutive optimizations in either an offline or an online game. We proved analytically that our mechanisms are truthful and cost-recovering. Through simulations, we demonstrated that our mechanisms also yield high utility compared with a regret-based state-of-the-art approach.

## 10. ACKNOWLEDGMENT

The astronomy simulation dataset was graciously supplied by T. Quinn, F. Governato, and S. Loebman of the UW Dept. of Astronomy. We also thank S. Loebman for providing the astronomy use-case and working with us on it. We also thank Nodira Khoussainova, Paraschos Koutris, Emad Soroush, and the anonymous reviewers for their comments on early drafts of this paper. This work is partially supported by NSF grant CCF-1047815 and Microsoft.